\documentclass[num-refs]{wiley-article}

\papertype{Original Article}
\paperfield{Journal Section}

\usepackage{epstopdf}
\usepackage{hyperref}
\usepackage{gensymb}
\usepackage{multirow}

\title{Real-time MRI-based fetal femur length measurement}

\author[1,2,3]{Johannes Barcsay*}
\author[4,5]{Sara Neves Silva*}
\author[2,3,4,5]{Jordina Aviles Verdera}
\author[4,5,6]{Charline Bradshaw}
\author[4,5]{Mary Rutherford}
\author[3]{Susanne Schulz-Heise+}
\author[2,3,4,5]{Jana Hutter+}

\affil[1]{Faculty of Computer Science, TH Deggendorf, Germany}
\affil[2]{Smart Imaging Lab, Radiological Institute, University Hospital Erlangen, Erlangen, Germany}
\affil[3]{Radiological Institute, University Hospital Erlangen, Erlangen, Germany}
\affil[4]{Early Life Imaging Department, School of Biomedical Engineering and Imaging Sciences, King’s College London, London, UK}
\affil[5]{Imaging Physics and Engineering Department, School of Biomedical Engineering and Imaging Sciences, King’s College London, London, UK}
\affil[6]{Department of Women and Children’s Health, King’s College London, London, UK}
\corremail{sara.neves\char`_silva@kcl.ac.uk}

\fundinginfo{* and + mark equal contribution. MRC grant [MR/X010007/1], Wellcome Trust fellowship [WT201526/Z/16/Z], UKRI FLF fellowship [MR/T018119/1], NIHR Advanced Fellowship [NIHR3016640], DFG Heisenberg [502024488], ERC StG [ERC101165242], High Tech Agenda of the Free State of Bavaria.
\vspace{5mm}}

\runningauthor{Barcsay et al.}

\begin{document}

\maketitle

\noindent \textbf{Purpose:} 
To develop and evaluate a real-time method for automatic planning and measurement of fetal femur length - an important indicator of antenatal growth - during MRI. While routinely assessed by ultrasound, MRI-based femur length measurements remain challenging due to bone-slice misalignment, fetal motion, and the need for manual assessment.
\\
\noindent \textbf{Methods:} 
A low-latency 3D U-Net was trained on 59 scans acquired at 0.55T (gestational age 18-40 weeks) to localise the proximal and distal endpoints of the femur. These coordinates were employed to automatically adapt a 30-second EPI sequence for in-plane femur coverage in real-time. Retrospective evaluation was performed in 72 scans (19-39 weeks, including 19 pathological cases), and real-time testing in 24 cases (17-39 weeks). Automated results were compared with manual expert annotations and, in 57/72 cases, with matched clinical ultrasound measurements. Precision was evaluated against inter-observer variability and analysed across gestational age and maternal BMI. Furthermore, bilateral femur length consistency was evaluated.
\\
\textbf{Results:} 
Retrospective analysis demonstrated a mean endpoint localisation error of 5.5 mm and femur length deviation of 3.0 mm. Among the 49/72 cases with both femora automatically extracted, the left-right difference was 2.8$\pm$2.7 mm. Precision was unaffected by maternal BMI (p$>$0.05), but correlated with gestational age (p$<$0.05). Real-time planning and assessment were successful in 22/24 cases, with a mean deviation of 4.2 mm.
\\
\noindent \textbf{Conclusions:} 
A real-time, automated method for MRI-based femur length measurement was developed, enabling precise, motion-robust, and operator-independent growth assessment, supporting future large-scale evaluation in growth-compromised pregnancies.

\newpage 
\section{Introduction}

\noindent Fetal magnetic resonance imaging (MRI) plays an increasingly important role in complementing ultrasound (US) \cite{levine2001ultrasound} for prenatal diagnostics, providing detailed anatomical visualisation and superior soft-tissue contrast, particularly in cases complicated by maternal obesity, oligohydramnios, or suboptimal fetal positioning \cite{Aboughalia2021Multimodality}. While the main clinical indications for fetal MRI remain pathologies of the central nervous system (CNS) \cite{prayer2023isuog}, a wide spectrum of genetic syndromes and skeletal dysplasias also manifest with abnormalities in fetal bone development, many of which have prognostic and therapeutic implications \cite{parilla2003antenatal}. Conditions such as thanatophoric dysplasia, osteogenesis imperfecta, campomelic dysplasia, and Jeune syndrome are characterised by shortened, bowed, or under-mineralised long bones \cite{Bondioni2017Comparative,poyner2013jeune}, often first suspected during routine prenatal screening \cite{Aboughalia2021Multimodality}. Moreover, chromosomal anomalies such as trisomy 21, 18, and 13 may present with shortened femur or humerus lengths as soft markers \cite{friebe2022femur}. Accurate prenatal assessment of long bones - particularly the femur - is therefore critical for diagnosis, monitoring, and counselling. It also contributes to gestational age estimation \cite{papageorghiou2016ultrasound}, fetal weight prediction, and the detection of growth restriction \cite{d2019midtrimester,Gaccioli2022}.

\noindent Ultrasound remains the first-line modality for fetal skeletal assessment, facilitated by its interactivity and real-time capabilities. However, increased depth of abdominal adipose tissue
and the more frequent presence of scar tissue from previous caesarean delivery possibly affect the quality of the acoustic window and subsequently impairs image quality \cite{tsai2015obesity}, while breech presentation is not only associated with increased obstetric risk but also with reduced accuracy of fetal weight estimation due to difficulties in obtaining reliable biometry, especially femur length \cite{melamed2011accuracy, toijonen2021risk}. MRI offers a valuable complement by providing a large field-of-view and consistent tissue contrast, independent of acoustic windows \cite{Nemec2011Fetal}. Yet, MRI-based femoral length assessment remains underutilised in clinical practice, primarily due to technical challenges in image planning and alignment \cite{davidson2021fetal, nogueira2018role}. MRI has intrinsic limitations for bone visualisation, including the absence of cortical bone signal due to its low proton density and rapid T2* decay, and its dependence on precise slice planning to capture long bones in-plane for quantitative measurement \cite{du2010qualitative}. Without targeted planning, the fetal femur may be visualised only partially or in oblique sections, precluding reliable length estimations \cite{Nemec2023Femur}. This is especially challenging in fetal imaging, where involuntary motion precludes long 3D acquisitions, making rapid 2D snapshot sequences the standard approach \cite{gholipour2014fetal}. The most commonly used sequence for fetal bone imaging, echo-planar imaging (EPI) with short echo times (TEs), is typically acquired in a 2D multi-slice approach to cover the region of interest \cite{Nemec2013Human}. Misalignment between slices often leads to oblique or fragmented bone representations, causing foreshortening artefacts and reducing measurement reliability \cite{uus2023retrospective}. As a result, the femoral diaphysis is frequently captured in cross-section or at an oblique angle, thus reducing diagnostic value. Accurate quantitative skeletal assessment thus hinges on precise plane prescription - a process that is currently manual, time-consuming, and highly operator-dependent. Furthermore, distortions inherent to EPI sequences affect the achievable measurement accuracy and further underscore the importance of positioning near the isocenter to minimise distortion artefacts. Recently, fetal MRI at 0.55T has been re-introduced \cite{ponrartana2023low,aviles2023reliability} as a strategy to mitigate such distortions. The wider bore enhances comfort, particularly accommodating the pregnant anatomy, while the lower scanner, maintenance, and infrastructure costs improve accessibility. In the dynamic fetal environment, manual planning remains inefficient, subject to inter-observer variability, and prone to suboptimal slice positioning \cite{matthew2018comparison}. These challenges highlight the need for intelligent, real-time planning strategies that adapt to fetal pose and optimise acquisition geometry during the MRI examination.\\

\noindent AI techniques have demonstrated strong potential for automation of fetal MRI workflows. Deep learning–based segmentation and motion-correction pipelines have enabled robust retrospective reconstruction and volumetric analysis 
of fetal structures \cite{uus2023retrospective,Pietsch2021}. Recent efforts have extended these capabilities to real-time applications: Hoffmann et al. \cite{hoffmann2021rapid} introduced a rapid head-pose detection framework for automatic slice alignment in fetal brain MRI, while Neves Silva et al. \cite{neves2024fully,silva2025automatic} proposed fully automated, real-time planning frameworks for both anatomical and flow imaging.\\

\noindent Building on these advancements, we propose a fully automatic, real-time planning and measurement pipeline for EPI-based imaging of fetal bones, with a particular focus on obtaining in-plane acquisitions of the fetal femur. Our framework integrates rapid anatomical localisation and AI-driven segmentation to identify long bones, estimate their orientations, and compute optimal scan planes in real-time. This approach is designed to be robust to fetal motion, reduce dependency on operator expertise, and ensure consistent femoral visualisation across subjects. In this work, we detail the development, implementation, and validation of this automated planning tool in a cohort of fetuses scanned on a clinical 0.55T MRI system. We assess its performance in terms of alignment accuracy, time efficiency, image quality, and reproducibility of femoral measurements. By enabling reliable, real-time planning of fetal skeletal acquisitions, this method has the potential to expand the clinical utility of MRI in assessing fetal growth and diagnosing skeletal pathologies.

\section{Methods}
\begin{figure}[!ht]
    \centering\includegraphics[width=0.95\textwidth,clip]{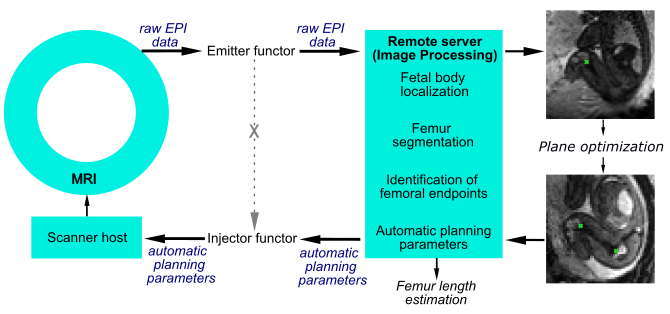}
\caption{Automated pipeline for MR-based femur length assessment. Step 1: initial scout acquisition and femur localisation; step 2: real-time computation of the optimal scan plane aligning the femur in-plane; step 3: adapted short-TE EPI scan with femur in-plane, enabling accurate length measurement.} \label{methodologypipeline}
\end{figure}

\noindent The complete real-time pipeline, graphically depicted in Figure \ref{methodologypipeline}, is described in the following sections. In step 1, the proximal and distal ends of the ossified shaft (diaphysis and metaphysis) are automatically identified; in step 2, the imaging plane parallel to the femur is calculated in real-time; in step 3, the in-plane scan is acquired, and the femur length (FL) measurement is automatically extracted.

\subsection{Step 1: Automatic identification of the proximal and distal femur end points}

\noindent A two-stage localisation and segmentation framework was implemented. First, the coarse region of interest is determined by segmenting the whole fetus using a pre-trained \cite{silva2025scanner,Payette2025} 3D nnU-Net \cite{isensee2021nnu}. Based on the estimated position of the fetus, the volume is cropped using a bounding box to ensure complete femur inclusion and rescaled to a uniform target size of 96x96x64 voxels. Second, femur segmentation within this sub-volume is performed using a 3D U-Net-based architecture derived from TransUNet \cite{chen2024transunet}. To enhance the robustness of the segmentation with respect to texture variability, a transformer-based refinement module incorporating multiple learnable femur representations is introduced in a subsequent step. Considering the initial femur probability map as input and region proposal, the most discriminative features of the U-Net decoder are sampled and integrated to refine segmentation and recover potentially missed structures. During training, a dynamically determined subset of labels is skeletonised to encourage a stronger emphasis on the central regions of the bone and further reduce the likelihood of fragmented segmentation. \\

\noindent Once the femur is localised and segmented, its endpoints and centre are extracted, transformed into scanner coordinates, and stored in a file for step 2, where these are applied to guide automated femur plane prescription. A separate file was also generated to record FL measurements performed on the initial scout scan, enabling later comparison with measurements obtained from the automated scan.

\paragraph{Fetal data acquisition}
\noindent In this study, 131 datasets were included - 59 for training and 72 for testing - acquired from pregnant women between 18 and 40 weeks of gestational age. All participants provided informed consent as part of ethically approved studies (MEERKAT [REC: 21/LO/0742], NANO [REC: 22/YH/0210] and MiBirth [REC: 23/LO/0685]). Fetal MRI scans were performed at St Thomas’ Hospital between 2022 and 2024 using a 0.55~T clinical MRI system (MAGNETOM Free.Max, Siemens Healthineers, Erlangen, Germany) and a combination of a 6-channel body coil and a 9-channel spine coil. Balanced steady-state free-precession (bSSFP) and multi-echo gradient-echo EPI (MEGE) sequences were employed in this study. All acquisition parameters are listed in Table \ref{tab:cohorts}. The fetal femur was manually annotated in all datasets.

\begin{table}[!ht] 
\centering 
\begin{tabular}{llcc} \hline
\multicolumn{1}{l}{\cellcolor[HTML]{FFFFFF}
\textbf{Dataset}}& 
\textbf{   Parameters} & 
\textbf{N} &
\textbf{Slices}
\\ 
\hline 
\rowcolor[HTML]{EFEFEF}
~Retrospective bSSFP & \small \begin{tabular} {l@{}l@{}}
Matrix = [288-480] × [288-576] mm \\
Resolution = 0.73 × 0.73 × [3-5] mm \\
TE = 4.0 ms \\
TR = 691.87 ms, total acquisition time = 17–48 s
\end{tabular} & 150 & 25–70 \\

\rowcolor[HTML]{FFFFFF}
\begin{tabular}{l@{}l@{}}
Retrospective MEGE EPI \\ \end{tabular} &  \small \begin{tabular}{l@{}l@{}}
0.55T Siemens MAGNETOM Free.Max \\
Matrix = 100x100–128x128 \\
Resolution 3.13–4.0 mm isotropic \\		
TE = [57, 152, 248, 344] ms \\
TR = 24.5 s, GRAPPA = 2 \end{tabular}
&  131& 62\\
\rowcolor[HTML]{EFEFEF}
 \begin{tabular} {l@{}l@{}}
Prospective EPI  \\ \small (scout scan) 
\end{tabular} & \small \begin{tabular}{l@{}c@{}}
0.55T Siemens MAGNETOM Free.Max \\
Matrix = 100x100–128x128 \\
Resolution 3.13–4.0 mm isotropic \\		
TE = 81 ms,  TR = 24.5 s (non-accelerated) \\
\end{tabular}
 & 24 &  62 \\

\rowcolor[HTML]{FFFFFF}
\begin{tabular} {l@{}l@{}}
Re-acquisition EPI \\ \small (in-plane femur)
\end{tabular} & \small \begin{tabular}{l@{}c@{}}
0.55T Siemens MAGNETOM Free.Max \\
Matrix = 100x100–128x128 \\
Resolution 3.13–4.0 mm isotropic \\		
TE = 57 ms, 10 slices, TR = 12.4 s, GRAPPA = 2 \end{tabular}
& 7 & 10 \\

\end{tabular} 

\caption{Information on acquisition parameters and number of datasets included in each of the cohorts employed in this study. bSSFP: balanced Steady State Free Precession; MEGE: Multi-Echo Gradient-Echo EPI.}
\label{tab:cohorts}
\end{table}

\subsection{Step 2: Planning of in-plane fetal femur planes}
\noindent A whole-uterus EPI sequence, previously modified to allow the acquisition of multiple echoes (MEGE), was further enabled to provide real-time detection of the fetal femur and computation of the coordinates required for automated plane prescription. The FIRE framework \cite{chow2021prototyping} was employed for real-time communication between the scanner and the AI networks. Processing was carried out on an external GPU-accelerated workstation (NVIDIA GeForce RTX 2080 Ti, NVIDIA Corporation), connected directly to the scanner computer. A subsequent EPI sequence (single echo) was further adapted to incorporate these coordinates for automatic planning of the fetal femur scan. The file generated in step 1 was transferred, in real-time, from the workstation to the scanner host computer to be accessible to the sequence, and the stored coordinates were applied as follows: the two endpoints of the segmented femur were used to compute the femur plane, while the centre point was used to define the slice stack centre and field-of-view. In parallel, the acquired image was processed with FIRE to extract real-time femur length measurements (see step 3). 

\noindent The self-prescribed scan retained all parameters of the above-described MEGE sequence, except that only 10 slices were acquired using a single TE (57~ms), resulting in a total run time of 4~s.

\subsection{Step 3: Automatic length measurement}

\begin{figure}[!ht]
    \centering
    \includegraphics[width=1\textwidth,clip]{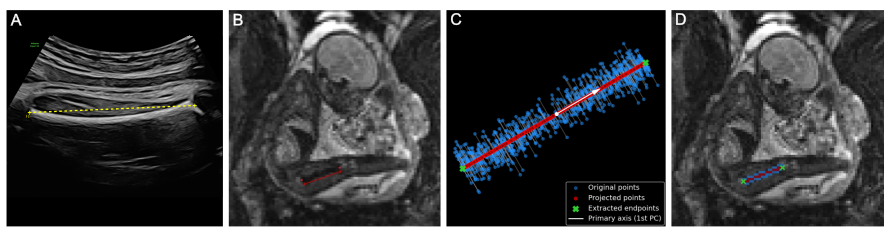}
    \caption{Illustration of the length measurement. (A) On ultrasound, the femur length is measured between the proximal and distal ends of the diaphysis. (B) In MRI, this corresponds to the hypo-intense ossified part of the bone. (C) Femoral landmarks are identified using Principal Component Analysis (PCA) on the generated network segmentation. The first principal component (white arrow) points in the direction of greatest variance and is used as the best-fit approximation of the anatomical axis. (D) The femur length was estimated as the distance between the points with the largest and smallest projection values (green cross).}
    \label{fig:figure5_pca}
\end{figure}

\noindent Figure \ref{fig:figure5_pca} depicts the appearance of the fetal femur on ultrasound (A) and MRI (B) and illustrates the automatic extraction of the femoral landmarks (C and D). Following previously established ultrasound guidelines, the femoral length is defined as the distance between the proximal and distal ends of the femoral diaphysis, including the metaphysis \cite{papageorghiouINTERGROWTHultrasound}. In EPI acquisitions, this corresponds to the hypo-intense part of the bone. Each femur label is considered a set of 3D coordinates, where each point corresponds to a voxel belonging to the ossified femoral shaft. Since the femur exhibits a predominantly straight and elongated morphology, the anatomical axis is approximated by a linear model obtained by Principal Component Analysis (PCA). The first principal component exhibits the direction of maximum variance in the data and thus aligns with the primary elongation of the femur. All points are orthogonally projected onto this principal axis, and the points with the minimum and maximum projection values are chosen as endpoints of the femur. Mimicking calliper-based measurements, the Euclidean distance between the extracted endpoints is set as the femur length (FL).

\subsection{Retrospective validation}
\noindent The method was retrospectively evaluated on a total of 72 datasets, with the scan setup as described in step 1. This cohort included scans acquired at gestational ages between 18 and 39 weeks (53 healthy, 19 non-healthy, including fetal growth restriction, short long bones, and multiple other pathologies).\\

\noindent Manual FL measurements were performed for both femur bones by two observers (M1, M2) to robustly estimate the error of the automatic method and assess inter-observer variability in manual and automatic measurements. The obtained FL was further compared to ultrasound measurements and biometry, available for 57 cases out of the 72 fetal MRI scans. The performance of the femur segmentation network was evaluated based on the overlap between manual and automatic segmentations using the Dice similarity coefficient, and contour alignment using 95\% Hausdorff distance (HD95).\\

\noindent The precision of the extracted landmarks was assessed by measuring the distance between automatic and manually determined endpoints, as well as by evaluating the accuracy of the resulting FL. Differences in FL were quantified using the mean absolute error (MAE) and mean absolute percentage error (MAPE) with respect to the manual measurements of M2. Measurement consistency was evaluated by comparing the obtained FL for both femur bones of the same fetus. Measurement error was further compared between second- ($\leq 27$~weeks) and third-trimester ($>27$ weeks) datasets. Automatic and manual length estimations were correlated with gestational age using Pearson correlation, and deviations were quantified. Kendall’s tau correlation analysis was performed to assess any potential correlations between MAPE and maternal body mass index (BMI). \\

\noindent Cases were divided into two groups based on the difference between MRI-derived and ultrasound-derived FL ($\leq 5\%$ and $> 5\%$). The number of slices covering the femur was quantified for each case, and group differences were statistically assessed using a two-sided t-test.

\subsection{Real-time prospective validation}
\noindent Similarly to the retrospective cohort, the 24 prospective datasets were acquired on a 0.55~T clinical MRI scanner (setup described above in step 1) with ultrasound measurements available for all cases (performed within 4 days of the fetal MRI examination). The complete pipeline was tested prospectively in real-time, starting with the initial localisation of the femur via a scout scan, followed by in-plane femur acquisition and subsequent FL computation.\\ 

\noindent For these 24 cases, accuracy was evaluated by comparing the real-time automatically generated segmentations to manual endpoint labels, extracted post-scan. Femur measurements were calculated manually and automatically, and compared with the ultrasound-based measurements.\\

\noindent Table \ref{tab:characteristics} provides a detailed overview of maternal characteristics and obstetric factors for the retrospective and prospective cohorts.\\

\begin{table}[!ht] 
\centering 
\begin{tabular}{l c c c c c }\hline 
\multicolumn{1}{c}{\cellcolor[HTML]{FFFFFF} 
\textbf{Cohort}} & 
\textbf{N} & 
\textbf{Gestational Age (weeks)} & 
\textbf {BMI} & 
\textbf{FL MRI (mm)} &
\textbf{FL US (mm)}
\\ 
\hline 
\rowcolor[HTML]{EFEFEF} 
Retrospective& 72 &  35.4 $\pm$ 4.8 & 31.2 $\pm$ 7.3 & 63.7 $\pm$ 10.3 & 70.1 $\pm$ 2.9\\ 
\rowcolor[HTML]{FFFFFF} ~~~ \textit{Healthy} 
& 53 &  36.5 $\pm$ 3.4 &  30.4 $\pm$ 5.9 & 66.7 $\pm$ 6.3 & 70.3 $\pm$ 2.4\\ 
\rowcolor[HTML]{FFFFFF} ~~~ \textit{Non-Healthy} 
& 19 &  32.3 $\pm$ 6.5 & 33.5 $\pm$ 9.9 & 55.8 $\pm$ 14.5 & 69.3 $\pm$ 4.1\\ 
\rowcolor[HTML]{EFEFEF} 
Prospective& 24 &  36.1 $\pm$ 5.3 & 27.8 $\pm$ 2.7 & 62.4 $\pm$ 11.7 & 66.9 $\pm$ 9.1\\ 
\end{tabular} 
\caption{Demographics of the training, retrospective and prospective cohorts. The given femur length measurements were obtained via manual measurement performed by fetal experts and obstetricians.} 
\label{tab:characteristics} 
\end{table} 

\section{Results}
\noindent The described pipeline was successfully implemented and evaluated in a total of 96 cases - 72 retrospective and 24 prospective cases.

\subsection{Retrospective validation}

\noindent Following successful fetal body detection and extraction of the relevant sub-volume in 71/72 cases, the femur was detected in 70 of these cases. In one data set in which the location of the fetal body failed, the extraction of the femur was still successful. Overall, femur detection was achieved in 71 out of 72 cases. Neither femur endpoint could be detected in one case, corresponding to a fetus of a gestational age of 18 weeks. This dataset, shown in supporting figure S1, was excluded from further analysis. In 49 out of 72 cases, both femur bones were detected. The complete pipeline - including body localisation, femur segmentation, and endpoint extraction - required an average of 6~s per volume (range 4.2-8.6 s) in offline testing mode. The femur segmentation achieved a Dice similarity coefficient of 0.66~$\pm$~0.1 and a HD95 of 5.0~$\pm$2.9~mm relative to manual reference segmentations. \\

\noindent The average spatial distance between manually and automatically identified femoral endpoints was 5.5~$\pm$~3.2~mm [range 0.0-20.0~mm], with the maximum distance observed for a case with poor bone contrast and image artefacts in the femoral region. The resulting MAE between manual and automatic annotations was 3.0~$\pm$~2.4~mm, which corresponds to a MAPE of 4.8~$\pm$~3.8$\%$. The mean length difference between the left and right femur was 2.7~$\pm$~2.8~mm (4.2~$\pm$~4.3$\%$) for manual measurements and 2.8~$\pm$~2.7~mm (4.4~$\pm$~4.4$\%$) for automatic FL estimations. The measurement error between the left and right femur was significantly correlated with the slice count difference ($r=.27, p < .05$) as illustrated in Figure S2 for manual measurements. \\

\begin{figure}[h!]
    \centering
    \includegraphics[width=1\linewidth]{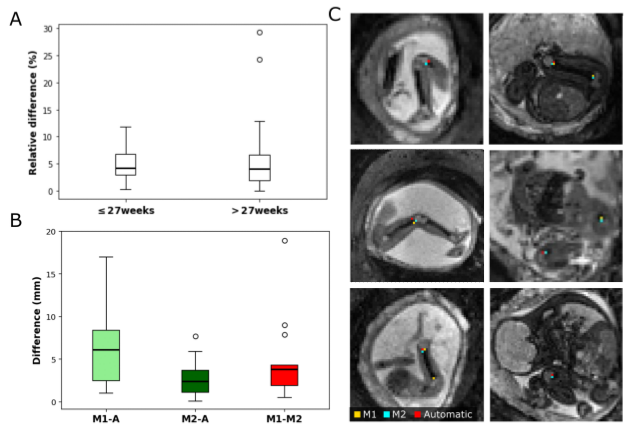}
    \caption{Evaluation of 72 retrospective cases. (A) Comparison of relative measurement errors in the second- and third-trimester datasets. (B) Difference in femur length measurements between manual annotations by two observers (M1, M2) and the automatic method (A). (C) Overlay of automatic and manual endpoints on the image acquired at the first echo time.}
    \label{fig:retrospective_results}
\end{figure}

\noindent The MAE of the cases acquired during the second trimester (n~=~8) was significantly lower compared to the third trimester (n~=~64) datasets ($p=.037$). However, there was no significant difference between both groups in terms of relative error ($p=.565$) as depicted in Figure \ref{fig:retrospective_results} A. In fetuses of gestational age $\leq$ 27 weeks (n~=~8), the average difference between manual and automatic measurements was 2.0~$\pm$~1.35~mm (5.2~$\pm$~3.5~\%), whereas in those $>$ 27 weeks (n~=~63), it was 3.2~$\pm$~2.9~mm (4.8~$\pm$~4.8~\%). No difference in measurement error was observed between healthy cases (4.8~$\pm$~3.9~\%) and those with detected pathologies (4.6~$\pm$~3.4~\%). Both manual and automatic measurements of FL showed a strong, significant relationship ($p<.001$) with gestational age, $r=.89$ and $r=.92$, respectively. The Kendall’s tau correlation analysis revealed no significant relationship between measurement error (MAPE) and maternal BMI, $\tau = -.02,~ p = .809$.

\noindent Figure \ref{fig:retrospective_results} B illustrates the results of inter-observer variability comparing the two manual observers, M1 and M2, with the automatically estimated lengths. Femur segmentations were performed by observer M2, and in all cases, manual and automatic measurements were derived from annotated endpoints. Absolute differences between M1 and automatic measurements ranged from 1.0 to 11.7~mm (mean:~6.2~mm), while differences between M2 and automatic measurements ranged from 0.5 to 7.8~mm (mean:~5.4~mm). Differences between the two manual observers ranged from 0.0 to 9.0~mm (mean: 4.9~mm). For six cases, the variability of endpoint annotations is depicted in \ref{fig:retrospective_results} C. \\

\noindent Manual measurements performed on the EPI acquisition were on average 2.6~$\pm$~3.9~mm (3.6~$\pm$~5.5~\%) smaller than corresponding ultrasound measurements, while the automatic estimations were 1.0~$\pm$~4.1~mm (1.4~$\pm$~6.1~\%) smaller. Cases in which the MRI-derived FL differed by more than 5\% from the ultrasound measurement showed a significantly greater number of slices spanning the femur (3.8 vs 7.3, $p=.017$) compared to cases with an error below 5\%.

\subsection{Prospective validation}

\begin{figure}[!ht]
    \centering
\includegraphics[width=0.95\textwidth,clip]{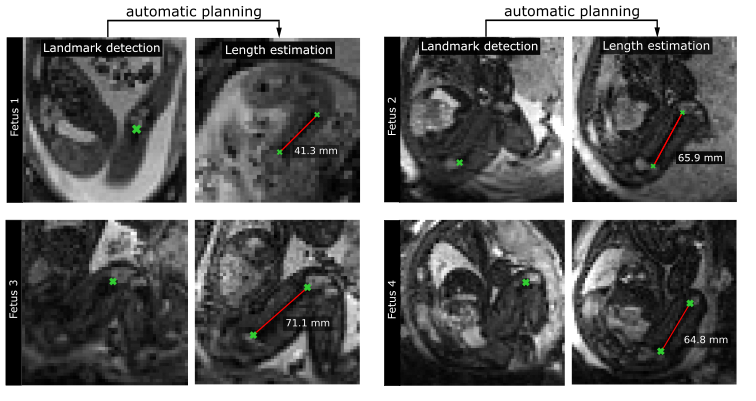}
\caption{Illustrative results of the complete pipeline in 4 prospectively acquired fetuses (23-39 weeks gestational age). (Step 1) Landmark detection was performed on the initial scout scan and enabled full coverage of the femur in plane during the (Step 2) re-planned acquisition.
} \label{results_prospective}
\end{figure}

\noindent The described pipeline was successfully implemented prospectively in 24 cases. Figure \ref{results_prospective} presents a comprehensive visualisation of the real-time results obtained for four example cases in the prospective validation cohort (23 and 39 weeks of gestational age), illustrating the EPI scan acquired in the coronal orientation of the uterus, the detected landmarks, the automatically re-planned acquisition, and the resulting FL measurement overlaid on the EPI image.\\

\noindent The detection of femoral landmarks and automatic planning of acquisition parameters took less than 6 seconds after completion of the first EPI sequence. On the initial EPI scan, the femur was successfully detected in all 24 cases and corresponding landmarks were extracted in real-time during the scan to automatically plan the following in-plane acquisition. Manual endpoint annotations were performed by M2 after the scans were completed. The complete elongation of the femur was captured in 22 cases, while automatically extracted endpoints were located along the femoral diaphysis rather than at the anatomical proximal and distal ends in two cases. Accordingly, the mean distance between automatic and manual endpoints on the out-of-plane scan was 10.0~$\pm$~20.1~mm, resulting in a MAE of 4.1~$\pm$~4.3$\%$ and a MAPE of 6.7~$\pm$~6.7$\%$ after exclusion of two segmentations covering less than 30$\%$ of the bone. Manual measurements of the left and right femur bones of the same fetus deviated on average by 5.5~mm $\pm$6.2~mm. Automatic right-left FL measurements were not performed, as in most cases, only one femur was fully captured within the field-of-view and free from artefacts. \\

\noindent Of the automatically planned in-plane scans performed, all but one captured the femur in plane with both ends clearly visible. The second acquisition resulted in an MAE of 1.3~$\pm$~0.4~mm and an MAPE of 2.3~$\pm$~1.0$\%$. The differences between manual and automatic measurements ranged between 0.7 and 1.8~mm, while automatic endpoints deviated on average 3.9~$\pm$~1.7~mm from the manual annotation. In-plane length measurements differed from the initial estimation by 2.3 $\pm$~0.9~mm for the automatic method and by 2.6~$\pm$~0.4~mm for manual measurements. The mean deviation from ultrasound biometry was 1.5~$\pm$~0.6~mm.\\

\begin{figure}[!ht]
    \centering
\includegraphics[width=0.95\textwidth,clip]{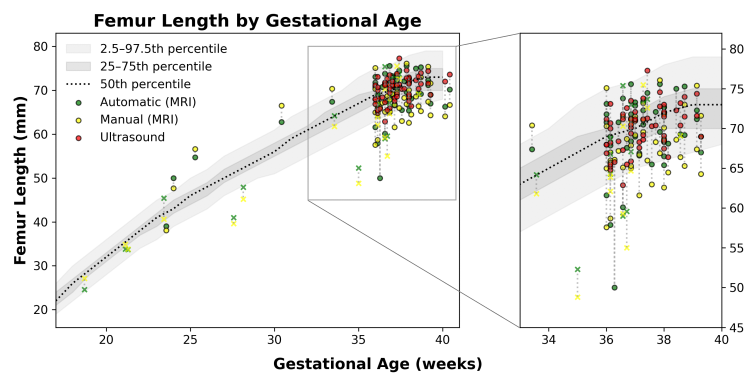}
\caption{Quantitative results shown for the entire retrospective cohort from 19 to 40 weeks of gestational age (left) and with a zoom into the last 5 weeks of pregnancy (right). The automatic MRI results are shown in green, the manual MRI results in yellow and the ultrasound comparison measurements in red for all cases where available. Cases with pathologies are indicated by a cross. The ultrasound-based percentile lines are highlighted in grey. } \label{fig:results_quant}
\end{figure}

\noindent Figure \ref{fig:results_quant} presents for the entire retrospective cohort MRI-based FL (manual and automatic) against gestational age, as well as the matched ultrasound results when available, stratified into healthy and pathological cases (marked with a cross) and overlaid on normative curves from large cohort ultrasound studies \cite{kiserud2017world}.\\

\begin{figure}[!ht]
    \centering
\includegraphics[width=0.95\textwidth,clip]{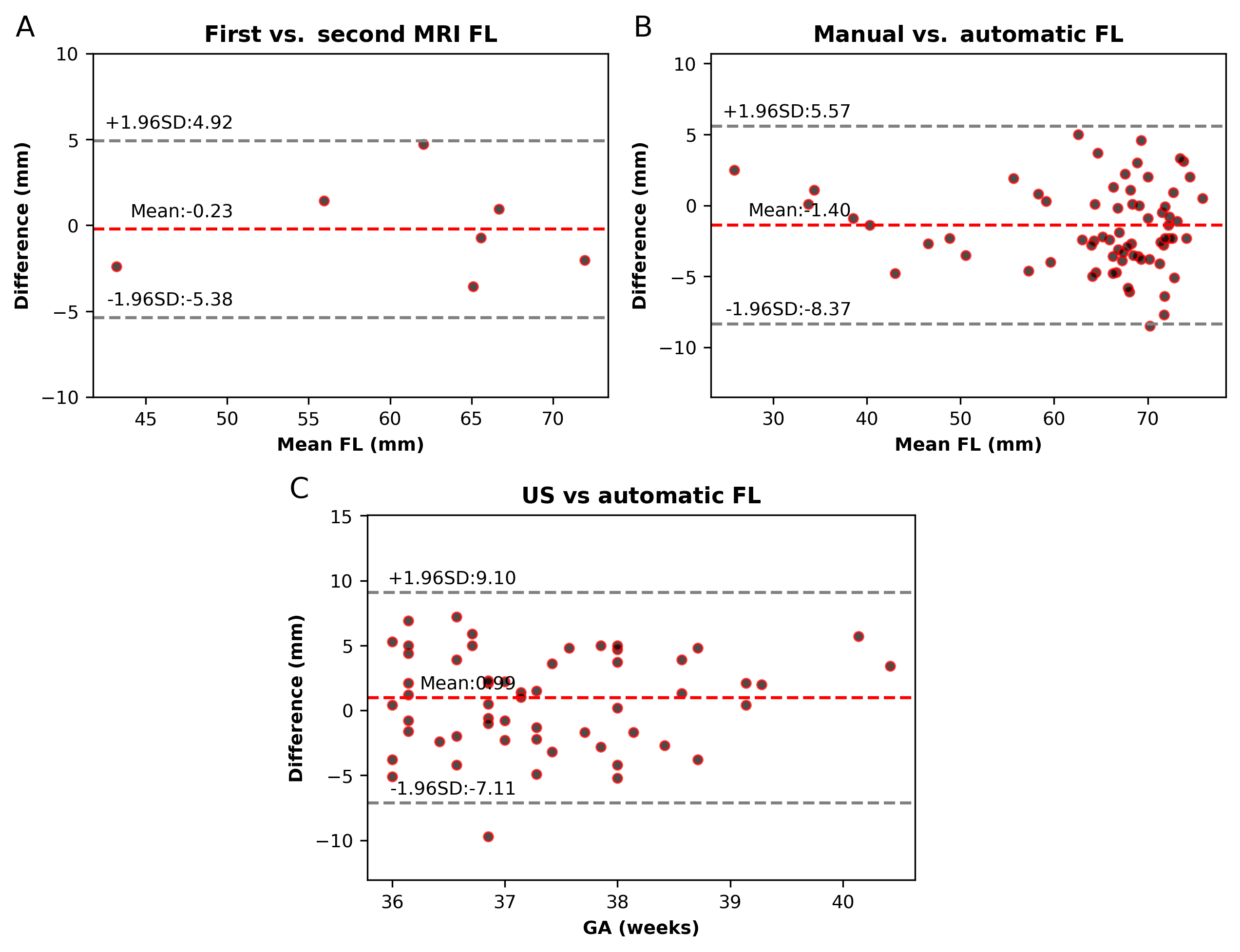}
\caption{Bland-Altman plots for robustness evaluation. (A) Differences between first and second automatic MRI-based FL measurements, (B) differences between automatic and manual MRI-based FL measurements, and (C) differences between automatic MRI and manual ultrasound measurements.} 
\label{fig:results_robust}
\end{figure}

\noindent The results of the robustness study are presented in Figure \ref{fig:results_robust}, with a mean difference in FL of 0.2~mm between the first and second real-time acquisitions illustrated in the Bland-Altman analysis (Fig. \ref{fig:results_robust}A). The mean difference between automatic and manual FL equals 1.4~mm (see Fig. \ref{fig:results_robust}B), and between ultrasound and MRI 1.0~mm (see Fig. \ref{fig:results_robust}C). The difference between ultrasound and MRI-based automatic FL correlates with size ($r=-.56, p<.05$) but not with gestational age ($p>.05$).  

\section{Discussion}

\noindent We presented a fully automatic real-time pipeline for quantitative fetal femur evaluation - consisting of fetal femur segmentation, landmark extraction, re-acquisition in the most suitable plane, and length estimation. The approach detects femoral endpoints in an initial scout scan and subsequently plans an in-plane acquisition in less than a minute, thus reducing operator workload and facilitating standardised MRI-based skeletal assessment. The MRI-based femur length could thereby be a substitute for ultrasound-based measurements in cases where precise length assessment in the latter is complicated. Although recent work typically acquired fetal bone scans at 1.5~T or 3~T \cite{Nemec2013Human}, this method was performed on a 0.55~T MRI scanner, where approximately 40$\%$ longer T2* relaxation times \cite{Campbell2019Opportunities} have the potential to capture the rapidly decaying signal from the bones more effectively. However, the methodology developed in this study is not limited to the application at 0.55~T.\\

\noindent Retrospective evaluation demonstrated reliable landmark detection with mean endpoint deviations of 5.5~mm (range 0.0-20.0~mm) and FL errors of 3.0~mm (range 0.0-16.1~mm) (4.9 $\%$), with the relative error in femur length remaining consistent across gestational age and comparable between healthy and pathological cases. Prospective testing confirmed the robustness of the achieved pipeline, as even partial detection of the femoral shaft is sufficient for accurate automated plane prescription: while scout scans were often affected by artefacts, the automatically planned acquisitions successfully captured the femur in plane. In-plane estimations achieved higher accuracy (MAPE: 2.3~$\%$) than out-of-plane measurements (4.9~$\%$). Notably, automatic measurements showed fewer implausible reference percentiles than manual annotations, underscoring their robustness despite these limitations. Manual annotation variability, however, remained high: small differences in calliper placement translated into several millimetres of discrepancy, particularly in oblique or cross-sectional views. 

\noindent Compared to previous work, the accuracy achieved falls within the reported range of MRI-based brain biometry \cite{Zalevskyi2025Advances, She2023Automatic}. The demonstrated in-plane results were comparable to ultrasound measurements, with mean FL errors of 0.9–2.7~mm reported \cite{Wang2019Joint, slimani2023fetal}. In line with the literature, performance remains constrained by the increased voxel size of the employed EPI sequences and susceptibility artefacts.

Common equations, such as proposed by Hadlock \cite{hadlock1985estimation}, estimate fetal weight using ultrasound-based measurements of the occipitofrontal diameter and abdominal circumference in addition to the femur length. Growing evidence suggests that MRI can provide more accurate estimations of fetal weight \cite{kacem2013fetal}. However, fetal weight estimation using MRI typically relies on volumetric approaches, which complicates direct comparison with classical methods. Previous studies have reported a high degree of agreement between manual measurements of standard biometry parameters in MRI and ultrasound \cite{matthew2018comparison} and particularly demonstrated the feasibility of the automatic assessment of cranial measurements in MRI \cite{avisdris2021automatic}. An extension of the proposed method to measurements of the fetal head and abdomen could offer an MRI-based alternative for fetal weight estimation and allow a comparison between volumetric and classical approaches for fetal weight estimation in MRI.

\noindent Key strengths of this study include the robustness of the proposed method, carefully designed with additional region-of-box detection and transformer-based steps to ensure reliable and generalizable performance for clinical applications. Even in cases where the entire femur was not fully captured, sufficient information was captured through partial bone fragments to enable subsequent planning and re-acquisition. The low latency achieved demonstrates the real-time feasibility of real-time implementation, as carefully demonstrated in the nested prospective study. Multi-observer assessments and comprehensive evaluation of all pipeline steps - from landmark detection to quantitative measurements - pave the way for clinical translation.\\

\noindent However, this study has several limitations. First, although the presented cohort is relatively large for a prospective fetal study, expanding it would allow more comprehensive testing of the robustness of automatic planning and provide a more reliable estimation of the accuracy of automatic femur length estimations, particularly in the re-planned acquisitions. The gestational age range, while spanning from 19 weeks to term, is skewed toward later pregnancy weeks due to the opportunistic recruitment through a study focusing on birth. Future studies could focus specifically on earlier stages of pregnancy. Furthermore, the current 3D approach faces challenges in robustly detecting the femur on the optimised acquisition. Visibility of the bone on two consecutive slices is required for reliable detection, and the choice of an appropriate slice thickness in the protocol is thus paramount.\\

\noindent Ultrasound biometry datasets were used as the reference standard in this study, a requirement due to the limited availability of MRI-based measurements, despite the widely acknowledged variability of ultrasound data. Furthermore, recent work also suggests a systematic bias, with the femur length in MRI EPI-based measurements typically shorter compared to ultrasound, and highlights the impact of operator experience on manual measurements \cite{matthew2018comparison}. These factors hinder further evaluation of the automatic method.\\

\section{Conclusion}
\noindent The successful integration of a complete pipeline for real-time, automatic femur length measurement in fetal MRI - demonstrated with robust and precise results across 96 participants - paves the way for future clinical integration and more individualised and accurate assessment of fetal growth. 

\section*{Acknowledgements}
The authors thank all pregnant women and their families for taking part in this study. The authors thank the research midwives and obstetric fellows for their invaluable efforts in recruiting and looking after the women in this study as well as perinatal radiographers for their involvement in the acquisition of these datasets. This work was supported by MRC funding [MR/X010007/1 and MR/W019469/1], UKRI DTP [2604718], DFG Heisenberg funding [502024488], an ERC StG [101165242] to JH, an NIHR Advanced Fellowship to LS [NIHR3016640] and by core funding from the Wellcome/EPSRC Centre for Medical Engineering [WT203148/Z/16/Z].


\newpage
\clearpage



\end{document}